# Model Vapor-Deposited Glasses: Growth Front and Composition Effects

*Ivan Lyubimov[*], M. D. Ediger[+], and Juan J. de Pablo[*]*

[*]Institute for Molecular Engineering, University of Chicago
5747 S. Ellis Avenue, Chicago, IL 60637, USA
[+]Department of Chemistry, University of Wisconsin – Madison
1101 University Avenue, Madison, WI 53706, USA

**Abstract**

A growing body of experimental work indicates that physical vapor deposition provides an effective route for preparation of stable glasses, whose properties correspond in some cases to those expected for glasses that have been aged for thousands of years. In this work, model binary glasses are prepared in a process inspired by physical vapor deposition, in which particles are sequentially added to the free surface of a growing film. The resulting glasses are shown to be more stable than those prepared by gradual cooling from the liquid phase. However, it is also shown that the composition of the resulting glass, which is difficult to control in physical vapor deposition simulations of thin films, plays a significant role on the physical characteristics of the material. That composition dependence leads to a re-evaluation of previous results from simulations of thinner films than those considered here, where the equivalent age of the corresponding glasses was overestimated. The simulations presented in this work, which correspond to films that are approximately 38 molecular diameters thick, also enable analysis of the devitrification mechanism of vapor-deposited glasses. Consistent with experiments, it is found that this mechanism consists of a mobility front that propagates from the free interface into the interior of the films. Eliminating surface mobility eliminates this route of transformation into the supercooled liquid.

A glass is typically prepared by cooling a liquid. Because glasses are out of equilibrium with respect to the liquid, the thermodynamic properties of a glass depend on the thermal history of the sample. The glass transition temperature, for example, typically decreases by a few degrees with every decade decrease of the cooling rate and glasses formed by slower cooling are generally stiffer and higher density. Glasses slowly evolve into structures that are deeper in the energy landscape in a process referred to as structural relaxation or ageing[1-3]. Recent experiments have shown that it is possible to create glasses with remarkable kinetic stability through the process of physical vapor deposition[4]. Such glasses exhibit higher density[4, 5], higher mechanical moduli[6], and lower enthalpy[4, 7-10] than any liquid-cooled glasses. These vapor-deposited glasses apparently correspond to the material that one would obtain by aging an ordinary glass over periods in the range of thousands of years or more[11].

Highly stable vapor-deposited glasses present a number of characteristics that make them potentially useful for applications, and it is therefore of interest to understand the origin of their unusual properties. While detailed studies have so far been limited to a few organic molecules, a number of intriguing questions have already emerged. These include whether stable glasses truly correspond to the state that ordinary glasses would reach after ageing for prolonged periods of time, or whether they constitute new amorphous states that cannot be accessed by cooling a liquid sample. Experimental studies of vapor-deposited glasses, for example, typically reveal structural anisotropy that is absent in ordinary glasses[12], and it is unclear whether one can create highly stable small-molecule vapor-deposited glasses whose structure is truly identical to that of the corresponding ordinary glass. Another important question concerns the extent to which stable glass mixtures can be formed by co-deposition of two molecules[13]; this might provide a promising avenue for discovery of new amorphous glassy materials with desirable properties.

Theoretical and computational studies of vapor-deposited glasses have been limited[14-17]. In recent work, we presented a method to create model vapor-deposited glasses in which one adopts a growth procedure that is inspired by that used in experiments[18, 19]. That method was used to prepare glasses of trehalose, a disaccharide of glucose, and glasses of a binary Lennard-Jones mixture[20] that has been used extensively in the past to study ordinary glasses. In both cases we were able to generate materials whose properties are consistent with those observed experimentally for several organic materials. In the case of trehalose, the simulated vapor-deposited glasses were shown to be structurally anisotropic. In the case of the binary Lennard-Jones glass, the materials were found to be isotropic.

Simulations of vapor-deposited glasses are highly computationally demanding. Our past computational studies of stable glasses were therefore limited to films of intermediate thickness, approximately 19 molecular layers or diameters. In this work, we consider binary Lennard-Jones vapor-deposited films that are much thicker than those studied in our original publication. With thicker samples at our disposal, we examine explicitly the effect of the substrate and the vacuum-film interface on the behavior of the film. We also examine the temporal relaxation of vapor-deposited films when the material is heated, and compare our results to recent experiments[21]. In agreement with experiment, we find that stable glasses transform into supercooled liquids by propagating fronts that can originate at free surfaces, and that the stability of such glasses can be enhanced by eliminating surface mobility.

The results presented in this work indicate that, consistent with observations from earlier simulations of smaller samples, it is possible to form highly stable vapor-deposited glasses of binary Lennard-Jones mixtures. Our new results, however, also reveal that the binary Lennard-Jones model is particularly sensitive to composition inhomogeneities introduced by the substrate and the free interface. Such inhomogeneities have a particularly strong effect on the energy and density of stable glasses, and require that we revisit our earlier interpretation of results for smaller samples. We do so by comparing the characteristics of vapor-deposited glasses to those of an ordinary glassy material prepared by heating and cooling a stable glass through the glass transition temperature. We find that vapor-deposited glasses have higher density and lower energy than ordinary glasses when compared at the same pressure and composition. We also find that vapor-deposited glasses correspond to the ordinary glasses that one would obtain by employing cooling rates that are approximately two to three orders of magnitude smaller than those typically accessible in simulations. This difference is significantly smaller than that anticipated on the basis of earlier comparisons to a bulk ordinary glass at a constant density and constant composition. However, the results reported here are comparable to results achieved with other computational approaches[22, 23] and are consistent with those observed in atomistic simulations of more realistic models[18].

**Model and Methods**

As described in previous work, we use a simulation procedure that attempts to mimic physical vapor deposition. The vapor-deposited glasses are grown by adding Lennard-Jones (LJ) particles, ten at a time, to the free interface of a growing film. The model considered here is based on that proposed by Kob and Andersen[24] and is the same as that employed in Ref.[19]. The substrate where the film is grown is maintained at temperature $T_s$ using a thermostat[25]. We use periodic boundary conditions in the directions parallel to the substrate, and the substrate consists of Lennard-Jones particles of diameter $\sigma_s=0.8$ and $\varepsilon_{ss}=0.6$. These particles are fixed in place using harmonic springs in a random arrangement at a density of $\rho_s=1.38$. New glass particles are introduced at random positions in the vicinity of the growing interface of the films, at a temperature T=1. The energy is then minimized using the FIRE algorithm[26]. After minimization, a molecular dynamics simulation is run using two thermostats: the substrate is maintained at $T_s$, and the newly introduced particles are equilibrated at T=1 for $10^5$ time steps. After that equilibration, the newly introduced particles are cooled down to the substrate temperature over a period of 7 $10^5$ steps (which corresponds to a cooling rate of approximately 3 $10^{-5}$). At that point, a new set of particles is introduced and the growth cycle is repeated. Unless otherwise indicated, the films considered here were grown to a total thickness of approximately 38 $\sigma_{AA}$. The dimensions of the films in the dimensions parallel to the substrate were fixed at 12.8 $\sigma_{AA}$. The total number of particles was approximately 6700 for full grown films. For all results presented in this work, the error bars corresponding to the vapor-deposited glasses and the ordinary glass films represent the standard deviation from 8 independent samples. For clarity, however, in Figures 5 and 6 results are shown for one representative configuration. For bulk samples, the error bars represent the standard deviation from five independent samples. The ``ordinary'' glasses considered here are prepared by heating the vapor-deposited glasses to a temperature well above $T_g$ (T*=0.55), and then cooling the films at a specified rate. The cooling rates considered here for ordinary films are in the range 3.33 $10^{-4}$ to 3.33 $10^{-6}$. All quantities are provided in reduced LJ units (t=$\sigma_{AA}(m/\varepsilon_{AA})^{1/2}$, T=$k_B T/\varepsilon_{AA}$, $\rho=\rho\ \sigma_{AA}^3$). The potential-energy landscape corresponding to the vapor-deposited and ordinary glass films is sampled by minimization of the potential energy. The thickness of the films is allowed to change during minimization, and the resulting inherent-structure configurations have different density than the original structures.

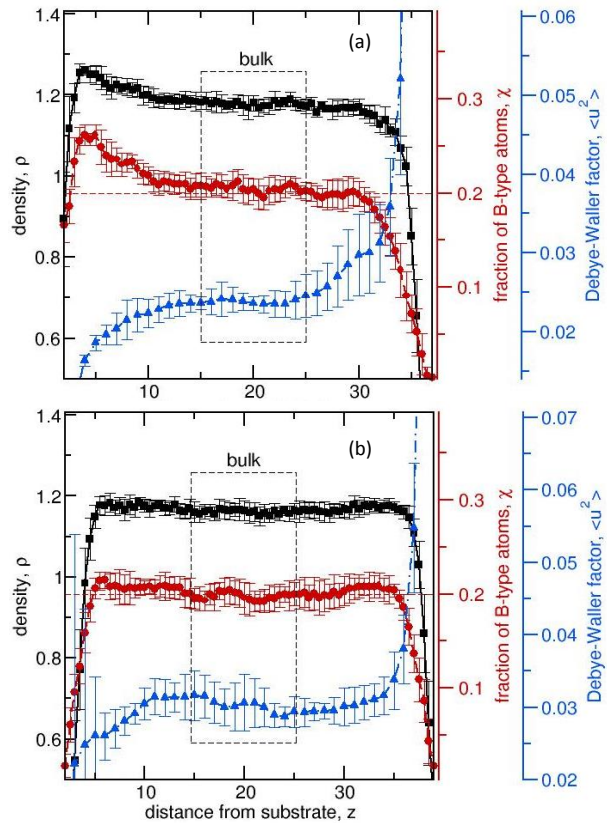

**Figure 1.** Density, composition, and Debye-Waller factors as a function of distance from substrate for (a) vapor-deposited glass films grown at a substrate temperature of $T_s$=0.3 and (b) ordinary glass films at T=0.3, created at a cooling rate of 3.33 10$^{-5}$.

**Results and Discussion**

As explained earlier, the vapor-deposited and ordinary glass films considered in this work are considerably larger than those employed in our previous work; we begin our discussion with a characterization of their macroscopic properties. Figure 1 (a) shows the density profiles of the vapor-deposited glasses grown at a substrate temperature $T_s$=0.3. The corresponding composition and Debye-Waller factor profiles at T=0.3 are also included in the figure. They represent an average from 8 independent samples. Figure 1 (b) shows the same information for the samples considered in Figure 1 (a), but after undergoing a heating and cooling cycle through the glass transition temperature, at a heating and cooling rate of 3.33 $10^{-5}$. For both vapor-deposited and ordinary glasses the density of the films is lower near the substrate or the free interface. In the middle of the films it is relatively uniform. The Debye-Waller factors follow the same trends as the density; near the substrate they decrease, but in the middle of the film, within the statistical uncertainty of our simulations they remain constant. In contrast, one can see that for vapor-deposited glasses (Figure 1 (a)) the concentration of B particles in the immediate vicinity of the substrate is significantly smaller than in the middle of the films; as the distance from the substrate increases, so does the concentration of B particles. It then goes through a maximum, before it decreases again to reach a plateau value after approximately 10 molecular layers. In this work, we define the "bulk" region of the film as that between $z$=15 to $z$=25. In that region, one can see that the density, the Debye-Waller factors, and the composition are relatively uniform and independent of the distance to the interfaces. Their respective average values for vapor-deposited glasses at T=0.3 are $\rho$=1.178(5), $x_B$=0.205(7), and $<u^2>$= 0.024(1). For ordinary glasses, the corresponding numbers are $\rho$=1.161(5), $x_B$=0.197(8), and $<u^2>$= 0.030(2). For the remainder of this paper, all results for ordinary glass and stable glass films refer to the "bulk" region of the films, except for Figure 6.

The potential energy as a function of temperature is shown in Figure 2 for the vapor-deposited glasses and for the ordinary glasses. The ordinary glasses were prepared at a cooling rate of 3.33 $10^{-5}$ and the energy during cooling is shown in the figure. One can infer a glass transition temperature of $T_g$=0.35 from the change in slope of the energy. For the vapor-deposited glasses, the abscissae correspond to the substrate temperature at which samples were generated. Below the ordinary glass transition temperature, the energy of the vapor-deposited glasses is significantly lower than that of the ordinary glasses. The difference becomes more pronounced as the substrate growth temperature decreases until the temperature approaches 0.15. Below $T_s$=0.125 the energy of the stable glass and the ordinary glass (data not shown) become comparable. The range of substrate temperatures in the vicinity of 0.3 represents an optimum for growth of vapor-deposited glasses, and it corresponds to approximately 80% of the glass transition temperature (see below). This observation is consistent with experiments with different substances, where it is found that the optimal substrate temperature for deposition of stable organic glasses is in the range of 0.8 to 0.85 $T_g$.[4, 7, 9, 10] The significance of this comparison is unclear, however, as the high cooling rates of simulations result in high $T_g$ values relative to experiments. One can extrapolate the equilibrium liquid potential energy to temperatures below $T_g$; the potential energy of vapor-deposited glasses lies on the extrapolated equilibrium liquid line down to a temperature slightly above T=0.3; below that temperature the vapor-deposited glass energy is higher than that of the extrapolated equilibrium supercooled liquid.

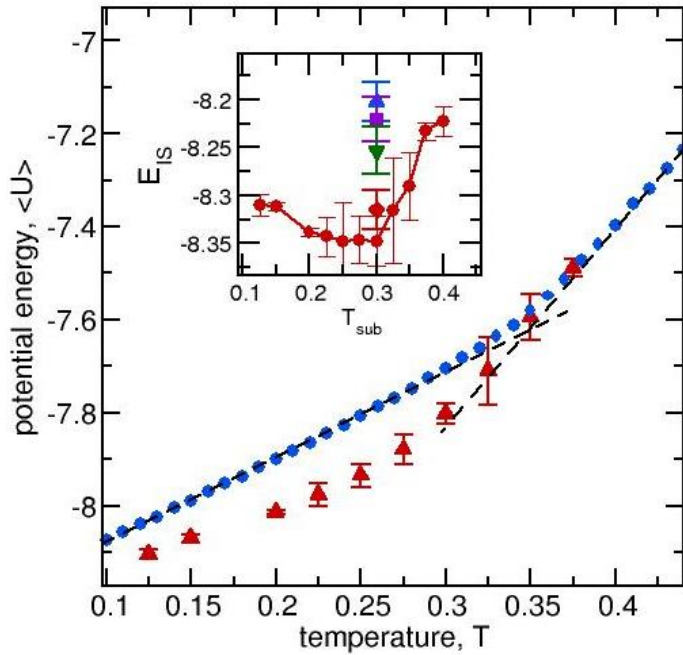

The inset in Figure 2 shows the corresponding inherent structure energy $E_{IS}$ for the vapor-deposited glasses (red symbols) and for ordinary glasses at three cooling rates (3.33 $10^{-4}$, 3.33 $10^{-5}$ and 3.33 $10^{-6}$ ). The red circles show results for samples of intermediate size, and the red diamond corresponds to results for large samples. One can see that the inherent structure energy of vapor-deposited glasses is relatively uniform for substrate temperatures in the range 0.2<$T_s$<0.3. At $T_s$=0.3, the vapor-deposited glasses exhibit a lower inherent structure energy than ordinary glasses. In our ordinary glass films, when the cooling rate changes from 3.33 $10^{-4}$ to 3.33 $10^{-5}$ to 3.33 $10^{-6}$, one observes an average change in $E_{IS}$ of approximately $\Delta E_{IS}$ = 0.025/decade, from -8.20(2) to -8.22(2) to -8.25(2). For vapor-deposited glasses the inherent structure energy is -8.32(2). That cooling rate dependence suggests that, in order to generate ordinary glasses with inherent structure energy comparable to that of the vapor-deposited glasses generated at $T_s$=0.3, the cooling rate would have to be approximately $10^{-9}$, i.e. approximately 2 to 3 orders of magnitude slower than that typically used in computer simulations of ordinary glasses.

**Figure 2.** Potential energy per particle for vapor-deposited glass films grown at substrate temperature $T_s$ (red) and ordinary glass films created at a cooling rate 3.33 $10^{-5}$ (blue). The ordinary glass transition temperature is estimated at $T_g$=0.35. The inset shows the inherent structure energy for stable glasses generated at different substrate temperatures (red circles and red diamond) and for ordinary glass films prepared at cooling rates 3.33 $10^{-4}$ (blue), 3.33 $10^{-5}$ (purple) and 3.33 $10^{-6}$ (green).

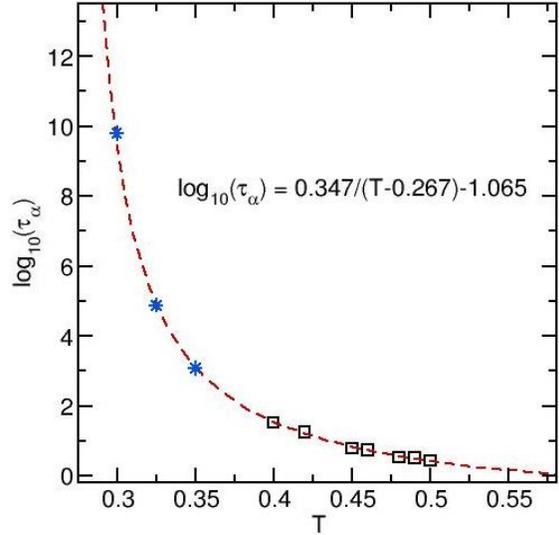

**Figure 3.** Open squares correspond to $\alpha$ relaxation time of bulk supercooled liquids at constant pressure (P=0, $x_B$=0.2). The blue symbols show the estimated relaxation times at a temperature of 0.3, 0.32 and 0.35, respectively.

An independent estimate of the age of vapor-deposited glasses can be inferred from the structural relaxation times of the material. Figure 3 shows the alpha relaxation times ($\tau_\alpha$) of bulk glasses generated at constant pressure. From Figure 2, we can estimate the fictive temperature of the glasses deposited at a substrate temperature of 0.3 is about 0.32.. From Figure 3, from the value of $\tau_\alpha$ at T=0.32, one can estimate a relaxation time for the vapor-deposited material that is approximately three orders of magnitude longer than that of the corresponding ordinary glass (with a fictive temperature of T=0.35). From a computational perspective, the ordinary glasses generated here at a cooling rate of 3.33 $10^{-6}$ required one week of CPU time; simulating glasses at cooling rates of $10^{-9}$ would require considerable more time than that available to us, and we are therefore unable to determine whether, after sufficient ageing, our ordinary glass films would adopt energies and structures similar to those of the vapor-deposited glasses.

Note that at T=0.3, upon minimization of the energy, the density of the inherent structure in the bulk region of ordinary films (prepared by heating the vapor-deposited glasses to T=0.55 and then cooling them at a rate of 3.33 $10^{-5}$) increases from 1.16 to 1.17, and that of the vapor-deposited glasses increases from 1.18 to 1.20. We attribute the difference between the densification of the ordinary and vapor-deposited glasses upon minimization to differences in the underlying structure of the two types of samples. To separate the effect of structure from that of the density, we also computed the inherent structure energy of a bulk glass under periodic boundary conditions with a composition of $x_B$=0.20 and a density of 1.20. The resulting energy is -8.260(3), which is higher than that observed in the vapor-deposited glass by approximately 0.06.

At this point it is important that we compare the results shown in Figure 2 for thick vapor-deposited glasses at constant pressure, to those presented in Ref.[19] for smaller samples. In Ref.[19], most of the samples were 19 layers thick, and the bulk region, identified by relying primarily on Debye-Waller factors, was closer to the substrate. That bulk region had a composition that was influenced by the substrate and was on average somewhat richer in B particles than the nominal composition of the model. The binary model considered here exhibits a pronounced effect on composition, particularly when examined at constant pressure, a point that has not been widely appreciated in the literature. This can be seen in Figure 4, which shows the potential energy and the inherent structure energy of ordinary bulk glasses for compositions ranging from $x_B$=0.12 to $x_B$=0.24. The solid circles were all generated at a constant density $\rho$=1.18 and at a temperature of T=0.3. One can see that, at constant density, the potential energy changes from -7.43 to -7.78 in that range. In the range from $x_B$=0.2 to 0.24, however, the potential energy changes weakly. Figure 4 also shows the potential energy of vapor-deposited glasses of different compositions (open squares), ranging from $x_B$=0.16 to $x_B$=0.24. In that range the energy changes appreciably, from -7.68 to -8.02. In particular, the bulk regions considered in Ref.[19] were on average in the range 0.22<$x_B$<0.24, and therefore exhibited energies that were systematically lower than those of samples with $x_B$=0.2.

The results reported here for vapor-deposited glasses correspond to constant pressure; the open triangles in Figure 4 correspond to the energy and inherent structure energy of ordinary glass films, prepared at a cooling rate of 3.33 $10^{-5}$. One can see that the energy of ordinary films exhibits the same composition dependence as that of the vapor-deposited glasses, but is always higher than that of the vapor-deposited material. For completeness, Figure 4 includes results for bulk glasses (with periodic boundary conditions) generated at constant pressure (P=0) and at a cooling rate of 3.33 $10^{-5}$. The average energy of the bulk glasses is comparable to that of the ordinary films considered here. The composition dependence of ordinary films and bulk samples is the same. The inherent structure energies of the bulk samples, however, are slightly higher than those of the ordinary films. We attribute this difference to the fact that when the energy of a thin film is minimized, the dimensions parallel to the substrate are kept constant, but the material is able to densify in the direction normal to the substrate. In contrast, the energy of the bulk samples was minimized by keeping all dimensions constant. Taken together, the results shown in Figure 4 indicate that the binary model considered in this work (and in Ref.[19]) is particularly susceptible to composition inhomogeneities, and a constant-density and constant composition reference state (i.e. the traditional Kob-Andersen model at $\rho$=1.18 and $x_B$=0.2), as was used in Ref.[19], is not appropriate. A better reference state to examine the stability of vapor-deposited glasses vis-à-vis that of ordinary glasses is that employed in Figures 1 (b) and 2, namely a glass film, generated at the same pressure, and subject to the same boundary conditions as the vapor-deposited glass.

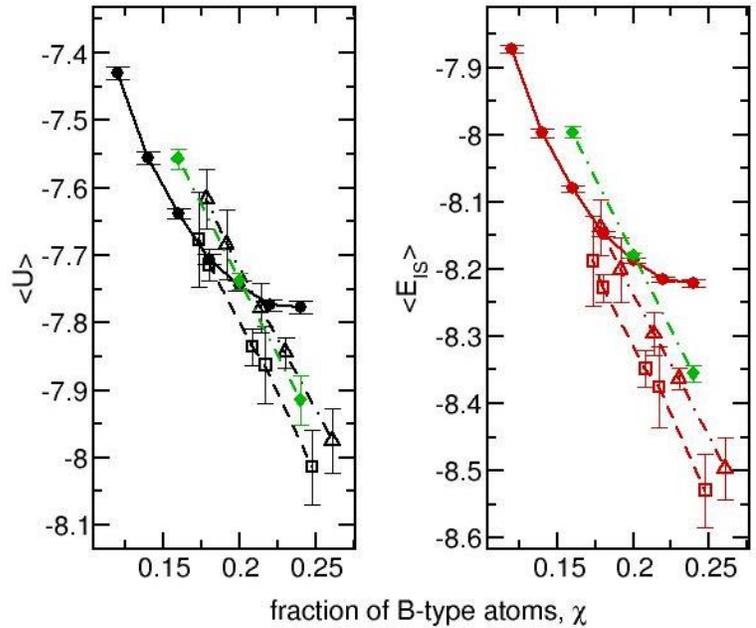

**Figure 4.** Potential energy (black) and inherent structure energy (red), per particle, for vapor-deposited glass films grown at a substrate temperature of $T_s$=0.3 (open squares), and for ordinary glass film prepared at a cooling rate of 3.33 $10^{-5}$ (open triangles). The solid black and red circles show results for ordinary glasses prepared under periodic boundary conditions at constant composition and constant density ($\rho$=1.18, solid symbols) also created at a cooling rate 3.33 $10^{-5}$. The solid green symbols show results for ordinary glasses prepared at constant pressure (P=0) at the same cooling rate.

Figure 5 shows the potential energy of the vapor-deposited glasses generated at $T_s=0.3$ as a function of temperature during heating, at a rate of 3.33 $10^{-5}$. For reference, the potential energy of an ordinary glass at the same heating rate is also shown in the figure. We refer to the "onset" as the temperature $T_o$ at which the glass is dislodged from its minimum in the energy landscape, and its potential energy suddenly increases towards its equilibrium liquid value. For ordinary glasses the onset temperature is comparable to $T_g$ in Figure 2. In contrast, for vapor-deposited glasses the onset temperature is $T_o=0.41$, approximately 14% higher than $T_g$. The devitrification process of Figure 5 is similar to that observed in experiments, where, depending on the molecule, the onset temperature for the most stable glasses are 6 - 10% above the glass transition[7, 27]. Given that vapor-deposited glasses have a lower potential energy and higher onset temperature than ordinary glasses, through the remainder of this manuscript we simply refer to them as "stable glasses".

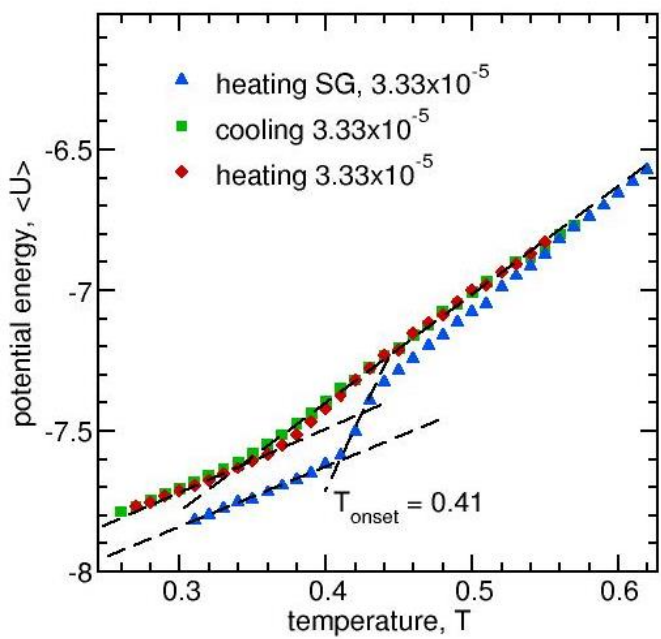

**Figure 5.** Potential energy per particle for vapor-deposited glass films grown at a substrate temperature of $T_s=0.3$ during heating at a rate of 3.33 $10^{-5}$ (blue symbols). The onset temperature for devitrification is estimated at $T_{onset}=0.41$. Results corresponding to an ordinary glass prepared at a cooling rate of 3.33 $10^{-5}$ (green) and then heated at the same rate (red) indicate that the onset temperature is comparable to the ordinary glass transition temperature ($T_g=0.35$).

Recent experiments have shown that when stable glass films are heated, devitrification of the glasses proceeds with a propagating mobility front moving into the interior of the material from the free interface[28, 29]. In contrast, when the stable glass films are "capped" with a thin layer of a different higher-$T_g$ material, the devitrification of the sample occurs more slowly and is initiated in the interior of the samples. In order to test for this behavior in our model stable glasses, a representative stable glass sample was brought to a temperature of T=0.42, which is slightly above the onset temperature. The sample was then held at constant temperature, and a molecular dynamics simulation was run to calculate the local Debye-Waller factors as a function time. The results are shown in Figure 6; one can see that, consistent with experimental observations, for stable glasses (Figure 6a) a front of mobility moves from the free interface into the interior of the samples as time proceeds. In contrast, Figure 6b shows results for a comparable isothermal simulation in which the material layers closest to the free interface were held at their original position, thereby mimicking the effect of a capping layer. That simulation was performed at constant density, but we have confirmed that the pressure remains constant throughout the time intervals shown in the figure. One can see that devitrification is now initiated well in the interior of the films, in agreement with the experimental observations of Sepulveda et al.[28] Note that devitrification now requires a longer time to proceed. Consistent with the experiments, this observation is interpreted to mean that once the surface-initiated transformation mechanism becomes inoperative, transformation must occur by a less efficient bulk mechanism. Figure 6c shows the corresponding results for an ordinary glass film. The sample is brought in one step to T=0.42, and the divitrification proceeds rapidly and uniformly across the film.

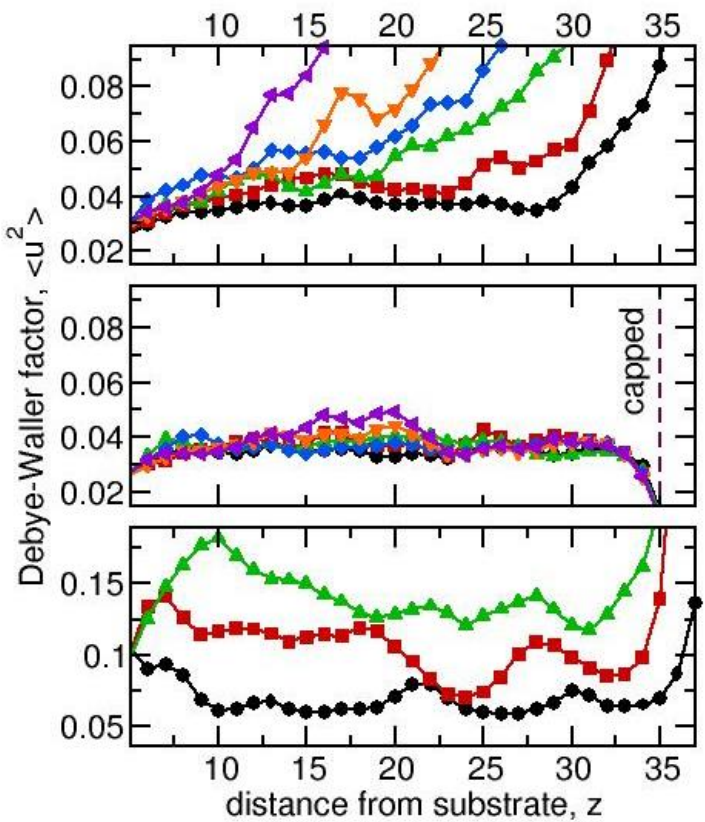

Our observations for how mobility fronts evolve in stable glasses supports the view advanced by Sepulveda et al.[28], in which particles of high mobility gradually release adjacent particles from the efficient packing of the stable glass. Since the free surface has a very high initial mobility, transformation from the stable glass into the supercooled liquid starts there and moves smoothly into the film interior. An interesting quantitative comparison can be made with these experiments. In Figure 6a, the growth front is observed to propagate a distance of about 20 $\sigma$ in a time interval of 56 $\tau_\alpha$, or about 0.3 particle diameters per $\tau_\alpha$. Experiments on two molecular glass formers show propagating front velocities of 0.1 to 0.01 molecular diameters per $\tau_\alpha$[28]. The observation that simulated stable glasses show front velocities within an order of magnitude of the experimental values suggests that further computational exploration of the transformation mechanism of stable glasses might be fruitful.

**Figure 6.** (a-b) Debye-Waller factors for vapor-deposited glass films grown at a substrate temperature of $T_s$=0.3 (black symbols) after bringing samples to T=0.42. Different colors correspond to different times during isothermal run (t=13 $\tau_\alpha$ -red, 24 $\tau_\alpha$ –green, 36 $\tau_\alpha$ -blue, 47 $\tau_\alpha$ -orange, and 58 $\tau_\alpha$ -purple). (a) Unrestricted stable-glass film. (b) Capped film, where the material layers located to the right of the dotted line were fixed in space. (c) Ordinary glass film (black symbols) prepared by cooling at a rate of 3.33 10$^{-5}$, bringing sample from T=0.3 to T=0.42 in one step, and then during isothermal run (t=13 $\tau_\alpha$ -red, 24 $\tau_\alpha$ – green).

**Conclusions**

The results presented in this work serve to confirm that one can generate stable glasses of a spherically symmetric model glass through a process that is reminiscent of experimental vapor deposition experiments. The resulting glasses are isotropic, and exhibit thermal and kinetic properties that are similar to those seen in experiments with small organic molecules. In particular, it is found that the density of stable glasses is approximately 1.5% higher than that of ordinary glasses at the same pressure and composition, the onset temperature is approximately 15% higher than the glass transition temperature, and the optimal substrate temperature for deposition is 20% lower than the glass transition temperature. The higher onset temperature is indicative of a material with a higher kinetic stability.

Recent experiments have shown that, in stable glasses, the divitrification of the material proceeds as a front moving from the free interface into the interior of the sample. When stable glass films are capped by a thin layer of a higher $T_g$ stable glass, devitrification is delayed and begins in the interior of the film. The same phenomenology is observed in the stable glasses generated here, with the occurrence of a distinct front propagating from the free surface into the film. In both experiments and in our simulations, divitrification of an ordinary glass proceeds homogeneously across the films and is more rapid.

The potential energy of our stable glasses at the optimal deposition temperature is considerably lower than that of the ordinary glass.  We estimate that the stable glasses presented here correspond to the ordinary glasses that one could generate at cooling rates that are about 2 to 3 orders of magnitude lower than those typically used in simulations of ordinary glasses. That figure is considerably smaller than the 19 orders of magnitude inferred in our previous work with smaller samples, in which the properties of stable glass films were compared to those for a reference material having constant density and a slightly different composition. Much of the unusually large effects reported earlier can be attributed to composition effects. The reference material considered here, namely an ordinary glass film at the same pressure and composition, provides a more appropriate framework in which to analyze stable glasses.  While the glasses prepared here are considerably more stable than those that can be prepared by cooling the liquid, they are considerably less stable than those prepared experimentally. Thus there remains considerable impetus for improved computational algorithms for generating low energy amorphous states.

**Acknowledgments**

The authors are grateful to Pawel Kosiatek, David Rodney and Jean-Louis Barrat for helpful discussions. This work is supported (IL, MDE and JDP) by grant NSF-DMR-1234320 .## 5. REFERENCES

1. C. A. Angell, Science **267** (5206), 1924-1935 (1995).
2. S. Sastry, P. G. Debenedetti and F. H. Stillinger, Nature **393** (6685), 554-557 (1998).
3. C. A. Angell, K. L. Ngai, G. B. McKenna, P. F. McMillan and S. W. Martin, Journal of Applied Physics **88** (6), 3113-3157 (2000).10